\documentclass[preprint,12pt]{elsarticle}
\usepackage{amssymb}
\usepackage{mathrsfs}
\usepackage{color}
\usepackage{amsmath}
\usepackage{amsthm}
\usepackage{graphicx}

\usepackage{txfonts} 

\newcommand{\tauin}{\tau^{(\mathrm{in})}}
\newcommand{\tauout}{\tau^{(\mathrm{out})}}
\newcommand{\Tin}{T^{(\mathrm{in})}}
\newcommand{\Tout}{T^{(\mathrm{out})}}
\newcommand{\Smain}{S^{(\mathrm{main})}}
\newcommand{\Sfast}{S^{(\mathrm{fast})}}
\newcommand{\argmin}{\mathrm{argmin}}
\newcommand{\R}{\mathbb{R}}
\newcommand{\N}{\mathbb{N}}

\renewcommand{\epsilon}{\varepsilon}

\newcommand{\VAR}{\mathtt{VAR}}

\newcommand{\COV}{\mathtt{COV}}
\renewcommand{\rho}{\varrho}
\renewcommand{\H}{\mathcal{H}}
\newcommand{\e}{\mathtt{e}}
\newcommand{\p}{\wp}
\newcommand{\pp}{\partial}
\newcommand{\K}{\mathcal{K}}
\renewcommand{\d}{\mathtt{d}}

\newcommand{\dw}{\mathtt{d}w}
\newcommand{\dt}{\mathtt{d}t}
\newcommand{\ds}{\mathtt{d}s}

\newcommand{\wopt}{w_{\mathrm{d}}}
\newcommand{\fit}{\mathtt{fit}}

\newcommand{\BE}{\begin{equation}}
\newcommand{\EE}{\end{equation}}

\journal{Physica A}

\begin{document}
\begin{frontmatter}

\title{Theoretical predictions for vehicular headways and their clusters}


\author{Milan Krb\'alek}

\address{Department of Mathematics, Faculty of Nuclear Sciences and Physical Engineering\\ Czech Technical University in Prague, Prague -- Czech Republic}

\begin{abstract}
This article presents a derivation of analytical predictions for steady-state distributions of netto time gaps among clusters of vehicles moving inside a traffic stream. Using the thermodynamic socio-physical traffic model with short-ranged repulsion between particles (originally introduced in [Physica A \textbf{333} (2004) 370]) we firstly derive the time-clearance distribution in the model. Consecutively, the statistical distributions for the so-called time multi-clearances are calculated by means of theory of functional convolutions. Moreover, all the theoretical surmises used during the above-mentioned calculations are proven by the statistical analysis of traffic data. The mathematical predictions acquired in this paper are thoroughly compared with relevant empirical quantities and discussed in the context of three-phase traffic theory.
\end{abstract}


\end{frontmatter}


\section{Introduction}\label{sec:prvni}

Explorations of traffic micro-quantities (i.e. quantities belonging to individual vehicles) and efforts to predict their statistics are as old as traffic research itself. Virtually, all important reviews on traffic science (for example  \cite{Review-Chowdhury, Review-Helbing, Review-Hoogendoorn}, or \cite{Review-Kerner}) try to explain at least some basic knowledge on statistical distributions of time-intervals or distance-gaps among moving cars. Although the theoretical prediction of time-evolution for headway-distributions is still extremely vague, some partial achievements have been reached in the last years (for example \cite{Buckley, Wagner, Knospe, Helbing_and_Krbalek, Kesting, Treiber_PRE, Magd, Surda, EPJ_Helbing, Li_and_Wang, Chen}, or \cite{Hrabak-ASEP}). We try to pick up the threads of those results and get closer to the heart of the matter.\\

This contribution is focused predominantly on the time-evolution of netto time intervals (so-called \emph{time-clearances}) between succeeding cars passing a given point (a traffic detector, typically) located at an expressway. Investigations of the traffic clearances are significantly advantageous (contrary to explorations of distance-gaps) because of their direct measurability. Indeed, the most of traffic detectors gauges the time of vehicle's passage directly, which means that those data do not show any systematic or mediated errors (provided the detector is not damaged). Except the time-clearances we concentrate our attention on the so-called \emph{multi-clearances}, i.e. cumulated time-clearances among $n$ consecutive cars. By analogy, such a quantity is (similarly to time-clearance) directly measurable, which opens a possibility for detailed statistical investigations of multi-clearance distributions with respect to the location of traffic ensemble in the fundamental diagram.\\

In fact, the empirical and/or theoretical investigations of traffic multi-headways are not sporadic in the physics of traffic. Some of the previous scientific works, e.g. \cite{Helbing-PRE, Red_cars, Cecile, Traffic_NV}, or \cite{Jin_and_Zhang}, briefly analyze the multi-clearance distributions (predominantly from the empirical point of view) or their statistical variances. The reasons for investigations of such a type are obvious, since deeper comprehension of changes in traffic microstructure will provide a more thorough insight into the convoluted traffic interactions.\\

\begin{figure}[htbp]
\begin{center}
\hspace*{-2mm}\includegraphics[width=7.5cm, angle=0]{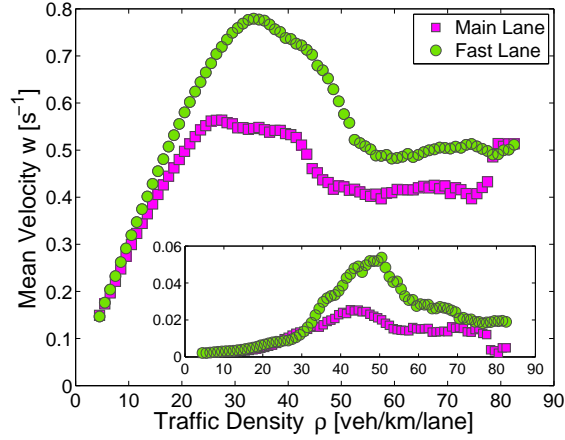}
\end{center}
\caption{Mean modified velocity w (and variance) as a function of traffic density. In the main part of the figure there is displayed the average value $\bar{w}$ which is drawn separately for the fast and main lanes. The fluctuations of velocity quantified by the velocity-variance $\VAR(W)=n^{-1}\sum_{k=1}^n (w_k-\bar{w})^2$ depending on traffic density are visualized in the inset.  \label{fig:Scaled-Velo}}
\end{figure}

\section{Preliminary evaluation of empirical observations}\label{sec:druha}

The vehicular-data records analyzed in this article are typically of the following types. The set
\BE \Tin_\ell=\bigl\{\tauin_{k\ell}\in\R^+|~k=1,2,\ldots,N\bigr\}, \quad \quad \bigl(\ell\in\{0,1,2\}\bigr) \label{Noname_01}\EE
includes the chronologically-ordered times when the front bumper of $k$th car has intersected the detector-line (located at $\ell$th lane of an expressway). The respective times $\tauout_{k\ell}$ when the rear bumper of $k$th car has intersected the detector-line are summarized in the set $\Tout_\ell.$ Analogously, the set of velocities of individual vehicles recorded by the detector is
\BE V_\ell=\bigl\{v_{k\ell}\in\R^+|~k=0,1,2,\ldots,N\bigr\}, \quad \quad \bigl(\ell\in\{0,1,2\}\bigr).\label{Noname_02}\EE
The lengths of vehicles are denoted as $d_{k\ell}$ and summarized in the set $D_\ell.$ We remark that the indices $\ell=0,1,2$ correspond to the slow, main, and overtaking lanes of a freeway, respectively. The slow lane is intended predominantly for long vehicles (trucks or lorries) and therefore the respective data will not be considered for this research. On contrary, the drivers use the third lane $(\ell=2)$ particularly if they are overtaking (and after overtaking-manoeuvre they turn back in the main lane), or if they are significantly faster that other cars. All of the above-mentioned quantities are (for purposes of this article) considered to be primary, which means that they are directly detecable by the traffic detectors.\\

Except the primary traffic quantities we now introduce some important secondary quantities, whose values are obtained vicariously, i.e. they are not included in the traffic-detector's records. The cardinal secondary traffic micro-quantities are \emph{the time-headways}
\BE z_{k\ell}:=\tauin_{k\ell}-\tauin_{(k-1),\ell}\label{Noname_03}\EE
and \emph{time-clearances}
\BE t_{k\ell}:=\tauin_{k\ell}-\tauout_{(k-1),\ell}.\label{Noname_04}\EE
Although both of them are not explicitly included in traffic-data files, their values are not burdened with any additional error. Above that, brutto space-gaps between successive vehicles (usually called as \emph{the distance-headways}) are traditionally approximated by the relation $s_{k\ell}:=v_{k\ell}~z_{k\ell}$ that presupposes the constant velocity $v(\tau)=v_{k\ell}$ during a time period when $\tau\in[\tauin_{(k-1),\ell},\tauin_{k\ell}].$ As well known, such a precondition is questionable, especially in the region of small traffic densities where the time-headways are too large. However, the influence of a possible error is of marginal importance, as apparent from the fact that the headway distributions analyzed in small-density regions do
not show any noticeable deviation from exponential distribution expected for infrequent events (see \cite{Krbalek_gas, Review-Helbing}, or \cite{Wagner}). Similarly, \emph{the distance-clearance} is calculated via $r_{k\ell}:=v_{k\ell}~t_{k\ell}$ and represents the estimated netto distance between $k$th car and its predecessor. Contrary to the time-headways (clearances) the distance-headways (clearances) are burdened by the systematic error that is discussed above. Such a error devalues the knowledge on distance-headway distributions and their evolution.\\

Denoting the sampling size by $m$ (in this study there is consistently considered $m=50$) and number of data-sets by $M_\ell$ (which therefore implies that $M_\ell=\lfloor N_\ell/m \rfloor$, where $N_\ell=\bigl|\bigl\{\tauin_{k\ell}|~\ell\text{ is fixed}\bigr\}\bigr|),$ one acquires the main data-samples
\begin{multline}\Smain_{j}=\Bigl\{\bigl(\tauin_{k\ell},\tauout_{k\ell},v_{k\ell},d_{k\ell}\bigr)\in\Tin_1\times\Tout_1\times V_1 \times D_1|~\\k=(j-1)m+1,(j-1)m+2,\ldots,jm~\wedge~\ell=1\Bigr\}, \label{Samples-main-lane}
\end{multline}
and the secondary data-samples
\begin{multline}\Sfast_{i}=\Bigl\{\bigl(\tauin_{k\ell},\tauout_{k\ell},v_{k\ell},d_{k\ell}\bigr)\in\Tin_2\times\Tout_2\times V_2 \times D_2|~\\
k=(i-1)m+1,(i-1)m+2,\ldots,im~\wedge~\ell=2\Bigr\},
\end{multline}
where $j=1,2,\ldots,M_1$ and $i=1,2,\ldots,M_2.$ For each data-sample $\Smain_{j}$ (or $\Sfast_{i}$ alternatively) we calculate the local flux
\BE J_j=\frac{m}{\tauout_{jm,\ell}-\tauin_{(j-1)m+1,\ell}} \label{Noname_05}\EE
and local average velocity $\bar{v}_j=m^{-1}\sum_{k=(j-1)m+1}^{jm} v_{k\ell}.$ The local density $\rho_j$ is then estimated (as suggested in \cite{Review-Helbing}) via the fluid-dynamic equation
\BE \rho_j=\frac{J_j}{\bar{v}_j}. \label{eq:fluid-dynamic equation} \EE
Such a expression is understood as one of the approximations suitable for estimation of vehicular density. We add that the incorrectness of the definition \eqref{eq:fluid-dynamic equation} is caused by the mixing of time and spatial averaging. Besides the macroscopic description of each sample $S_j$ we now introduce the mean time-clearance (or  distance-clearance) by means of definitions
\BE \bar{t}_j=\frac{1}{m}\sum_{k=(j-1)m+1}^{jm} t_{k\ell}, \quad \quad \bar{r}_j=\frac{1}{m}\sum_{k=(j-1)m+1}^{jm} r_{k\ell},\label{Noname_06}\EE
respectively.\\

\begin{figure}[htbp]
\begin{center}
\hspace*{-2mm}\includegraphics[width=7.5cm, angle=0]{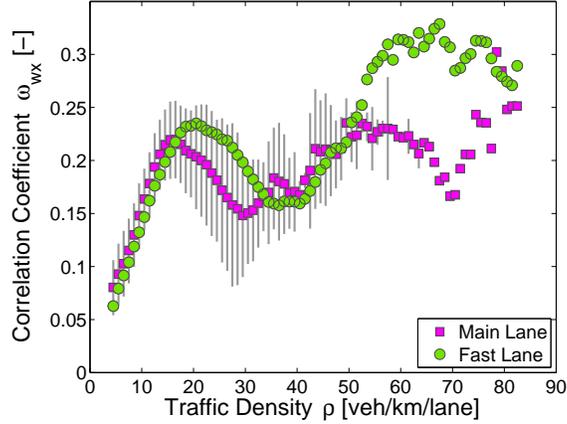}
\end{center}
\caption{Correlation coefficient  $\omega_{wx}$ as a function of traffic density. The value $\omega_{wx}$ quantifies a statistical correlation between the modified vehicular velocities $w_{k\ell}$ and scaled gaps $x_{k\ell}$ to the preceding car (in the regions of fixed traffic densities). The squares represent cars moving in the main lane, whereas circles correspond to the fast-lane cars. Note that the vehicles in the fast lane show the stronger correlations than others. Such a tendency is accented in the regions where the mental strain of drivers is stronger (i.e. in the regions of over-saturations). Vertical abscissae demonstrate a degree of fluctuations (quantified by the standard deviation) in $\omega_{wx}$ for main-lane data analyzed from different data sources. \label{fig:Corr-Coeff}}
\end{figure}

Since aiming to investigate the essential properties of micro-distributions (i.e. statistical distributions of microscopic traffic quantities) we eliminate (in the next part of this text) the global trends in those distributions. Predominantly, we eliminate the changes of the average headways caused by the varying traffic density. Such an approach corresponds to the technique examined in the articles (\cite{Helbing_and_Krbalek, Krbalek_gas, Kesting, Traffic_NV}, or \cite{Hrabak-ASEP}), and allows more sophisticated comparison for different traffic regimes or for traffic data originated from different countries. The procedure of the headway re-scaling (see the text below) represents in fact the trivial variant of the so-called \emph{Savitzky--Golay smoothing filter}  (for details please see the Ref. \cite{Savitzky-Golay}) applied to the matrix spectra in the Random Matrix Theory (see \cite{Mehta}), for example. Thus, we define the \emph{scaled distance-clearances} (for the sample $S_j$) as
\BE x_{k\ell}=\frac{r_{k\ell}}{\bar{r}_j} \quad \text{for all} \quad k\in\bigl\{(j-1)m+1,(j-1)m+2,\ldots,jm\bigr\} \label{Noname_07}\EE
and the \emph{modified velocities} (representing a local traffic flow, in fact) as
\BE w_{k\ell}=\frac{v_{k\ell}}{\bar{r}_j} \quad \text{for all} \quad k\in\bigl\{(j-1)m+1,(j-1)m+2,\ldots,jm\bigr\}.\label{Noname_08}\EE
It implies that the mean clearance in each sample is re-scaled to the unit. In the Fig. \ref{fig:Scaled-Velo} we demonstrate the changes of averages of the modified velocities $w_{k\ell}.$\\

For intentions of analytical calculations executed in the section \ref{sec:treti} it is now necessary to verify a statistical independence (or negligible dependency) between individual spatial clearances and car velocities. The standard way how to inspect this property is to investigate the associated correlation coefficient. Denoting $X=\{x_1,x_2,\ldots,x_n\}$ and $W=\{w_1,w_2,\ldots,w_n\}$ the sets of scaled distance-clearances or modified velocities respectively, one can calculated the Pearson's correlation coefficient using the relation
\BE \omega_{wx}= \frac{\COV(X,W)}{\VAR(X)\cdot\VAR(W)} =\frac{\sum_{k=1}^n (x_k-1)(w_k-\bar{w})}{\sqrt{\sum_{k=1}^n (x_k-1)^2}~\sqrt{\sum_{k=1}^n (w_k-\bar{w})^2}}, \label{Noname_09}\EE
where $\bar{w}=\sum_{k=1}^n w_k/n.$ Such a coefficient $\omega_{wx}\in[-1,1]$ represents a measure for stochastic dependency between $X$ a $W,$ so that the values being close to zero indicate a weak correlation.\\

Dividing the complete data-sample into the sub-samples of almost-constant traffic density we demonstrate in the Fig. \ref{fig:Corr-Coeff} that coefficient $\omega_{wx}$ is localized in the range $[0.05;0.30]$ and the corresponding values depend on a traffic phase (in the three-phase-traffic interpretation) and freeway lane. However, in all density-intervals the Pearson's correlation is restrained, which means that the hypothesis on a weak dependence between individual clearances and velocities seems to be legitimate.

\section{Mathematical derivation of time-clearance distribution}\label{sec:treti}

As introduced in \cite{Helbing_and_Krbalek}, justified in \cite{EPJ_Helbing}, and elaborated in \cite{Krbalek_gas}, one of the possible ways how to acquire theoretical predictions for distributions of empirical distance-clearances is to use the socio-physical traffic model whose elements are repulsed by the short-ranged power-law forces and randomized by the socio-physical noise, which is supposed to be of thermal-like nature. The influence of such a thermodynamical component to the model can be increased/reduced by the socio-physical coefficient $\beta\in[0,\infty)$ reflecting a mental strain under which the drivers are during a given traffic situation. Specifically, in the free traffic regimes (where the psychological pressure caused by the traffic situation is weak) the parameter $\beta$ is almost zero. On contrary, for the congested phase (where interactions among drivers are reinforced) the mental strain coefficient $\beta$ is large. For the detailed changes of $\beta$ one can inspect the articles \cite{Helbing_and_Krbalek, Krbalek_gas}, or \cite{Traffic_NV}.\\

To be specific, we consider dimensionless particles moving along a ring whose velocities are $w_1,w_2,\ldots,w_N$ and mutual distances (between subsequent particles) are $x_1,x_2,\ldots,x_N.$ Introducing the short-ranged repulsive potential
\BE U(x_1,x_2,\ldots,x_N)=\sum_{k=1}^N \frac{1}{x_k} \label{potencial} \EE
and socio-physical hamiltonian (see \cite{Krbalek_gas})
\BE \H(w_1,w_2,\ldots,w_N,x_1,x_2,\ldots,x_N)=\frac{1}{2}\sum_{k=1}^N \bigl(w_k-\wopt\bigr)^2 + U(x_1,x_2,\ldots,x_N) \label{hamiltonianek} \EE
one can derive (for details please see \cite{Krbalek_gas}) that velocities of such a ensemble (analyzed in the steady state) are gaussian-distributed, i.e. the associated probability density reads
\BE q(w)=\frac{1}{\sqrt{2\pi}\sigma} \e^{-\frac{(w-\wopt)^2}{2\sigma^2}}, \label{velocity-distribution} \EE
where $\sigma^2$ is the second statistical moment (variance) and $\wopt$ is the optimal velocity of drivers. By analogy, in the articles \cite{Krbalek_gas, Traffic_NV} there has been deduced that the scaled distance-clearance distribution reads as
\BE \p(r)=A~ \Theta(r)~\e^{-\beta/r}\e^{-Dr},
\label{p_beta} \EE
where $\Theta(x)$ is the Heaviside's step-function and the normalization factors are
\BE D \approx \beta+\frac{3-\e^{-\sqrt{\beta}}}{2}, \label{becko} \EE
\BE A^{-1}=2\sqrt{\frac{\beta}{D}}\K_1\bigl(2\sqrt{D\beta}\bigr).\label{acko} \EE
As verified in \cite{Helbing_and_Krbalek, Krbalek_gas, Red_cars} the one-parametric distribution-family \eqref{p_beta} is in an satisfactory agreement with the clearance distribution observed in the real-road data. We add that the noise-parameter $\beta$ is related to the corresponding traffic density $\rho.$\\

\begin{figure}[htbp]
\begin{center}
\hspace*{-2mm}\includegraphics[width=7.5cm, angle=0]{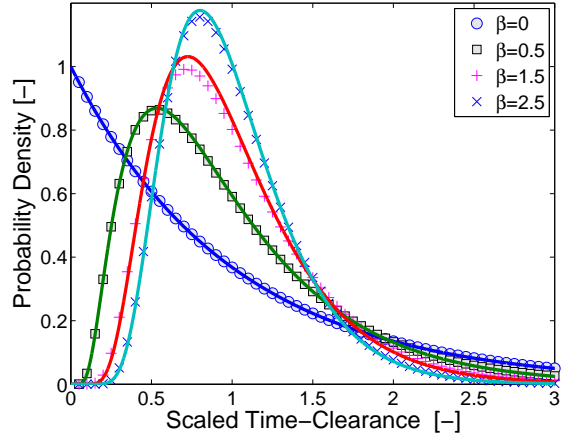}
\end{center}
\caption{Time-clearance distribution (analytically derived). The signs represent the values of function \eqref{TC-prvni-aproxka} calculated for $\beta\in\{0,1/2,3/2,5/2\},$ whereas the curves display the zero-approximation \eqref{TC-distribution} for the same $\beta.$ Small discrepancies between signs and corresponding curves are caused by fact that statistical variance $\sigma^2$ of the modified velocities (measured on freeways) is very low. Phenomenologically, this relation can be estimated (for the main-lane data) by the inequality $\sigma^2 \leq (45\beta/55)^4\exp[-45\beta/14].$ \label{fig:TC-distributiuon-general}}
\end{figure}

With the help of formulas \eqref{velocity-distribution} and \eqref{p_beta} one can mathematically derive the analytical form of probability density $\eta(t)$ for clear time-intervals between two consecutive vehicles (particles of a model). Since the joint probability density for distance and velocity can be (under the condition on weak distance-vs-speed dependency -- see the Fig. \ref{fig:Corr-Coeff}) predicted as $g(w,x)=q(w)\p(x),$ the joint probability density for time gaps and velocities reads therefore as $h(w,t)=wq(w)\p(wt).$ Then the time-clearance distribution represents a marginal density
\BE \eta(t)=\int_\R h(w,t)\,\dw = \int_\R wq(w)\p(wt)\,\dw.\label{Noname_10}\EE
After expanding a function $f(w)=w\p(wt)$ into the
Taylor's series about the optimal velocity $\wopt$ we acquire
\BE f(w)=f(\wopt)+\sum_{\ell=1}^\infty \frac{1}{\ell!} \frac{\d^\ell f}{\d w^\ell}(\wopt)(w-\wopt)^\ell,\label{Noname_11}\EE
where
\BE\frac{\d^\ell f}{\d w^\ell}=\frac{\pp^\ell \p}{\pp (wt)^\ell}t^\ell w +\ell \frac{\pp^{\ell-1} \p}{\pp (wt)^{\ell-1}}t^{\ell-1} \quad (\ell\in\N).\label{Noname_12}\EE
Hence
\begin{multline}
\eta(t)=\wopt\p(\wopt t)+\sum_{\ell=1}^\infty \frac{1}{\ell!}\frac{\pp^\ell \p}{\pp (wt)^\ell}(\wopt t) t^\ell \wopt \int_\R (w-\wopt)^\ell q(v)\,\dw~+\\
+ \sum_{\ell=1}^\infty \frac{1}{\ell!}\frac{\pp^{\ell-1} \p}{\pp (wt)^{\ell-1}}(\wopt t) t^{\ell-1} \int_\R (w-\wopt)^\ell q(w)\,\dw~=\\
=\wopt\p(\wopt t)+\sum_{\ell=1}^\infty \frac{\mu_\ell}{\ell!} \left(\frac{\pp^\ell \p}{\pp (wt)^\ell}(\wopt t) t^\ell \wopt  + \frac{\pp^{\ell-1} \p}{\pp (wt)^{\ell-1}}(\wopt t) t^{\ell-1} \right),
\end{multline}
where $\mu_\ell=\int_\R (w-\wopt)^\ell q(w)\,\dw$ is $\ell-$th central
statistical moment with three prerogatived cases $\mu_0=1,$
$\mu_1=0,$ and $\mu_2=\sigma^2.$ The latter represents a
statistical variance (see also \eqref{velocity-distribution}). As well known the odd central statistical moments (associated to the Gauss distribution) are zero and even moments comply with equalities $\mu_{2\ell}=\sigma^{2\ell}(2\ell-1)!!,$ which leads to the general formula
\BE \eta(t)= \wopt\p(\wopt t)+\sum_{\ell=1}^\infty \frac{\sigma^{2\ell}}{(2\ell)!!} \left(\frac{\pp^{2\ell} \p}{\pp (wt)^{2\ell}}(\wopt t) t^{2\ell} \wopt  + \frac{\pp^{2\ell-1} \p}{\pp (wt)^{2\ell-1}}(\wopt t) t^{2\ell-1} \right). \label{TH-general} \EE
Owing to the facts $\int_\R \wopt\p(\wopt t)\,\dt=1$ and
\BE \int_\R \left(\frac{\pp^{2\ell} \p}{\pp (wt)^{2\ell}}(\wopt t) t^{2\ell} \wopt  + \frac{\pp^{2\ell-1} \p}{\pp (wt)^{2\ell-1}}(\wopt t) t^{2\ell-1} \right)\dt=0,\label{Noname_13}\EE
the TC-distribution \eqref{TH-general} is normalized correctly. In a certain sense (as probably understandable from the two previous equations) the first factor in \eqref{TH-general} forms a leading term of TC-distribution, whereas the sum in \eqref{TH-general} represents a perturbation term only. Above that, the weight of the $\ell$th perturbation summands decreases with a factor $\sigma^{2\ell}/(2\ell)!!.$ For practical applications (especially for traffic applications) it seems therefore to be coherent using an approximate expansion
\begin{equation}
\eta(t) \approx \wopt\p(\wopt t) + \frac{\sigma^2}{2} \left(\frac{\pp^2 \p}{\pp (wt)^2}(\wopt t) t^2 \wopt + 2 \frac{\pp \p}{\pp (wt)}(\wopt t) t\right). \label{TC-obecna}
\end{equation}
Such an approximation is legitimate because of extremely low value of velocity variance mainly (see the inset in the Fig. \ref{fig:Scaled-Velo}). Note that the factors $\sigma^{2\ell}/(2\ell)!!$ measured in real-road data rapidly vanishes since the maximal value of velocity variance is $\approx 0.02$ in the main lane or $\approx 0.06$ in the fast lane. The expansion \eqref{TC-obecna} can be replaced (after the scaling procedure) by the formula
\begin{equation}
\eta(t) \approx \p(t) + \sigma^2\p(t)\left(\frac{\beta^2}{2t^2}+\frac{D^2t^2}{2}-D(t+\beta)\right). \label{TC-prvni-aproxka}
\end{equation}
The influence of the low velocity-variance to TC-distribution is demonstrated in the Fig. \ref{fig:TC-distributiuon-general} where the curves \eqref{TC-prvni-aproxka} are compared to the zero-approximation
\BE \eta(t) \approx A~ \Theta(t)~\e^{-\frac{\beta}{t}}\e^{-Dt}, \label{TC-distribution} \EE
while the relations (\ref{becko}),\,(\ref{acko}) hold true. We remind that the normalization and re-scaling conditions $\int_\R \eta(t)\,\dt= \int_\R t\,\eta(t)\,\dt=1$ have been applied.

\section{Mathematical derivation of multi-clearance distribution}\label{sec:ctvrta}

\begin{figure}[htbp]
\begin{center}
\hspace*{-2mm}\includegraphics[width=7.5cm, angle=0]{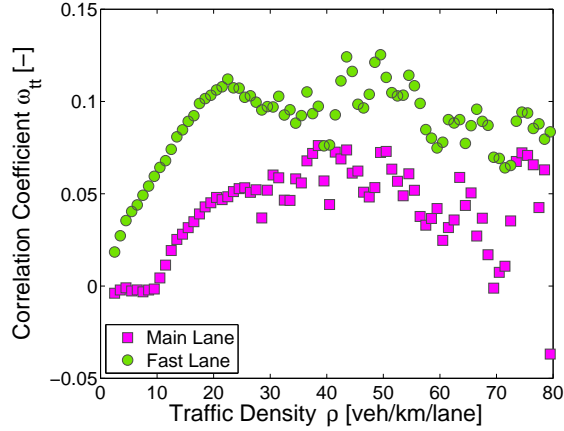}
\end{center}
\caption{Graphical representation of correlation analysis of successive time-clearances. The values $\omega_{tt}=\omega_{tt}(\varrho)$ describes the standard correlation coefficient calculated for pairs of succeeding time-clearances extracted from traffic data in various density regions.   \label{fig:Corr-Coeff-Headways}}
\end{figure}

Now, knowing the one parameter family of the probability densities
\eqref{TC-prvni-aproxka} or their zero-approximations \eqref{TC-distribution} one can derive an analytical prediction for the so-called \emph{$n$th multi-clearance distribution} $\tau_n(t)$ which represents the probability density for a clear time gap $t$ among
$n+2$ neighboring particles. Therefore, the main goal of this section is to quantify (by means of theory of functional convolutions) the probability that time-period between two following instants (the first one: the back bumper of the $k$th vehicle is leaving the detector; the second one: the front bumper of the $(k+n+1)$th vehicle is intersecting the detector line) is ranging in the interval $[t,t+\dt).$ Using this notation we find that the probability density for the time-clearance between two succeeding
cars (derived in the previous section) is $\eta(t)=\eta_0(t).$ Regarding the clearances as independent (which is demonstrated in the figure \ref{fig:Corr-Coeff-Headways}) the $n$th probability density $\eta_n(t)$ can be calculated via recurrent formula
\BE \eta_n(t)=\eta_{n-1}(t)~\star~\eta_0(t),\label{Noname_14}\EE
where symbol $\star$ represents a convolution of the two probabilities, i.e.
\BE \eta_n(t)=\int_{\R} \eta_{n-1}(s)\eta_0(t-s)\,\ds. \label{Noname_15}\EE

As a first resort, we will concentrate our endeavor to studying the low-density traffic regimes where the locations of cars are independent and therefore the traffic flow corresponds (from mathematical point of view) to the Poisson process of uncorrelated events. As published in \cite{Helbing_and_Krbalek,Krbalek_gas,Traffic_NV} such a situation is characterized by the negligible value of socio-physical coefficient $\beta$ reflecting a driver's mental strain. Thus, the associated clearance distribution reads $\eta(t)=\Theta(t)\e^{-t}$ and sequentially the multi-clearance distribution for free traffic looks like
\BE \eta_n(t)=\Theta(t)\,\e^{-t}\int_0^{t}\int_0^{s_1}\int_0^{s_2}\ldots\int_0^{s_{n-2}}\int_0^{s_{n-1}}\ds_{n}\ds_{n-1}\ldots\ds_2\ds_1 = \Theta(t)\frac{t^n}{n!} \e^{-t}. \label{multi-poisson}\EE
As a second resort, we aim to derive a general formula, i.e. formula for non-zero $\beta.$ For these purposes we use the zero-approximation \eqref{TC-distribution}. Applying the method of mathematical induction and an approximation of the function
\BE g_n(t,s)=\e^{-\beta\bigl(\frac{n^2}{s}+\frac{1}{t-s}\bigr)}\approx \e^{-\frac{\beta}{t}(n+1)^2} \label{saddle-point-approx} \EE
in the saddle point one can obtain
\begin{multline}
\eta_n(t)=\Theta(t)\int_0^t A_{n-1}A s^{n-1}\e^{-\beta\frac{n^2}{s}}\e^{-Ds}\e^{-\frac{\beta}{t-s}}\e^{-D(t-s)}\,\ds~=\\
=\Theta(t)A_{n-1}A \e^{-Dt}\int_0^t s^{n-1}g_n(t,s)\,\ds~\approx\\
\approx \Theta(t)A_{n-1}A  \e^{-\frac{\beta}{t}(n+1)^2}\e^{-Dt} \int_0^t s^{n-1}\,\ds \approx \Theta(t)A_{n-1}A n^{-1}  t^n \e^{-\frac{\beta}{t}(n+1)^2}\e^{-Dt}.
\end{multline}
Hence
\BE \eta_n(t)\approx \Theta(t)A_n t^n\e^{-\frac{\beta}{t}(n+1)^2}~\e^{-Dt}, \label{nta-clearancka}\EE
where (after applying the re-normalization procedure)
\BE A_n^{-1}=2\left(\sqrt{\frac{\beta}{D}}(n+1)\right)^{n+1}\K_{n+1}\bigl(2(n+1)\sqrt{D\beta}\bigr). \label{Noname_25}\EE
This fixes the proper normalization $\int_\R \eta_n(t)\,\dt=1.$ In
addition to that the mean $n$th spacing equals to
\BE \int_\R t\,\eta_n(t)\,\dt=n+1. \label{Noname_16}\EE
Note, that \eqref{nta-clearancka} holds true also for the limiting case $\beta=0.$ Really, with help of formulas $\lim_{x\rightarrow 0} x^{n+1}\K_{n+1}(x)=(2n)!!$ and $\lim_{\beta \rightarrow 0_+} D(\beta)=1$ we easily deduce that
\BE \lim_{\beta \rightarrow 0_+} 2\left(\sqrt{\frac{\beta}{D}}(n+1)\right)^{n+1}\K_{n+1}\bigl(2(n+1)\sqrt{D\beta}\bigr) = n!.\label{Noname_17}\EE
Therefore it holds $A_n=1/n!,$ which is in a full consonance with the relation \eqref{multi-poisson}.
Thus, the relation \eqref{nta-clearancka} constitutes a zero-approximation for the distribution of time multi-clearances. Since we have supposed (in the previous deductions) that the variance of modified velocities is negligible (in a local sense), the form of the distribution \eqref{nta-clearancka} is a direct consequence of the distribution for spatial clearances and the fact, that all cars have practically the same velocity (in a local sense, again).\\

Because the empirical measurements show low (but not negligible) variances in modified velocities (see the inset in the Fig. \ref{fig:Scaled-Velo}), it seems more realistic to derive the time multi-clearance distributions under the conditions $\sigma^2>0$ and $\sigma^{2n}\approx 0$ for $n=2,3,4,\ldots.$ Thus, we suppose that clear time-intervals among succeeding cars are distributed according the rule \eqref{TC-prvni-aproxka}. The detailed analysis of such a relations vindicates that the dominating term in the last summand of \eqref{TC-prvni-aproxka} is $\sigma^2Dt\p(t).$ Hence, we surmise that the time clearance follows the law
\BE \eta_0^{(\sigma)}(t)\approx \p(t)-\sigma^2Dt\p(t). \label{Noname_18}\EE
Here we remark (for mathematical correctness) that this function has to be understood as an approximative probability density, since the perturbation term $\sigma^2Dt\p(t)$ causes that $\eta_0^{(\sigma)}(t)$ does not fulfil the exact mathematical definition.\\

Using the method of mathematical induction we prove below that (under the previous surmises) the multi-clearance distribution of order $n$ reads
\BE \eta_n^{(\sigma)}(t)\approx \Theta(t)\frac{A^{n+1}}{n!} t^n \e^{-(n+1)^2\frac{\beta}{t}}\e^{-Dt}\bigl(1-\sigma^2Dt\bigr). \label{TC-sigma-aproximace}\EE
For completeness, we remark that in the following mathematical operations
\begin{multline}
\eta_{n+1}^{(\sigma)}(t) = \eta_n^{(\sigma)}(t) \star  \eta_0^{(\sigma)}(t) \approx \Theta(t)\frac{A^{n+2}}{n!}\e^{-Dt} \int_0^t \e^{-(n+1)^2\frac{\beta}{s}}s^{n} \e^{-\frac{\beta}{t-s}}~\ds~+\\
+D^2\sigma^4\Theta(t)\frac{A^{n+2}}{n!}\e^{-Dt} \int_0^t \e^{-(n+1)^2\frac{\beta}{s}}(t-s)s^{n+1}\e^{-\frac{\beta}{t-s}}~\ds~-\\
- D\sigma^2\Theta(t)\frac{A^{n+2}}{n!}\e^{-Dt} \int_0^t \e^{-(n+1)^2\frac{\beta}{s}}s^{n+1}\e^{-\frac{\beta}{t-s}}~\ds~-\\
- D\sigma^2\Theta(t)\frac{A^{n+2}}{n!}\e^{-Dt} \int_0^t \e^{-(n+1)^2\frac{\beta}{s}}(t-s)s^{n}\e^{-\frac{\beta}{t-s}}~\ds~=\\
= \Theta(t)\frac{A^{n+2}}{(n+1)!} t^{n+1} \e^{-(n+2)^2\frac{\beta}{t}}\e^{-Dt}\bigl(1-\sigma^2Dt\bigr)
\end{multline}
there have been used the approximation \eqref{saddle-point-approx} and the surmise $\sigma^4 \approx 0.$

\section{Empirical multi-clearance distribution vs. analytical prediction}\label{sec:pata}

In this section we will balance the theoretical prognoses deduced in the sections \ref{sec:treti} and \ref{sec:ctvrta} against the empirical multi-clearance distributions of freeway data analyzed in the section \ref{sec:druha} (two-lane freeway A9 - Netherland, two-lane freeway D1 - Czech Republic). In the following part of the text we will consider the multi-clearances (here: cumulated clearances between $n=5$ or $n=8$ succeeding cars)
\BE \varkappa_i^{(j)}=\sum_{k=(j-1)m+i}^{(j-1)m+i+n} \frac{t_{k\ell}}{\bar{t}_j}, \quad (\ell=1,~j=1,2,\ldots,M_1,~i=1,2,\ldots,m-n) \label{Noname_19}\EE
enumerated for data sets \eqref{Samples-main-lane}. These multi-clearances are associated with the specific traffic density through the formula \eqref{eq:fluid-dynamic equation} and re-scaled so that the average multi-clearance (in the given data-samples $\Smain_{j}$) is equal to $n+1.$ Now, the multi-clearances $\varkappa_i^{(j)}$ represent a statistical realization of random variable $t$ considered in the section \ref{sec:ctvrta} and can be therefore confronted with the theoretical probabilities.\\

\begin{figure}[htbp]
\begin{center}
\hspace*{-2mm}\includegraphics[width=10cm, angle=0]{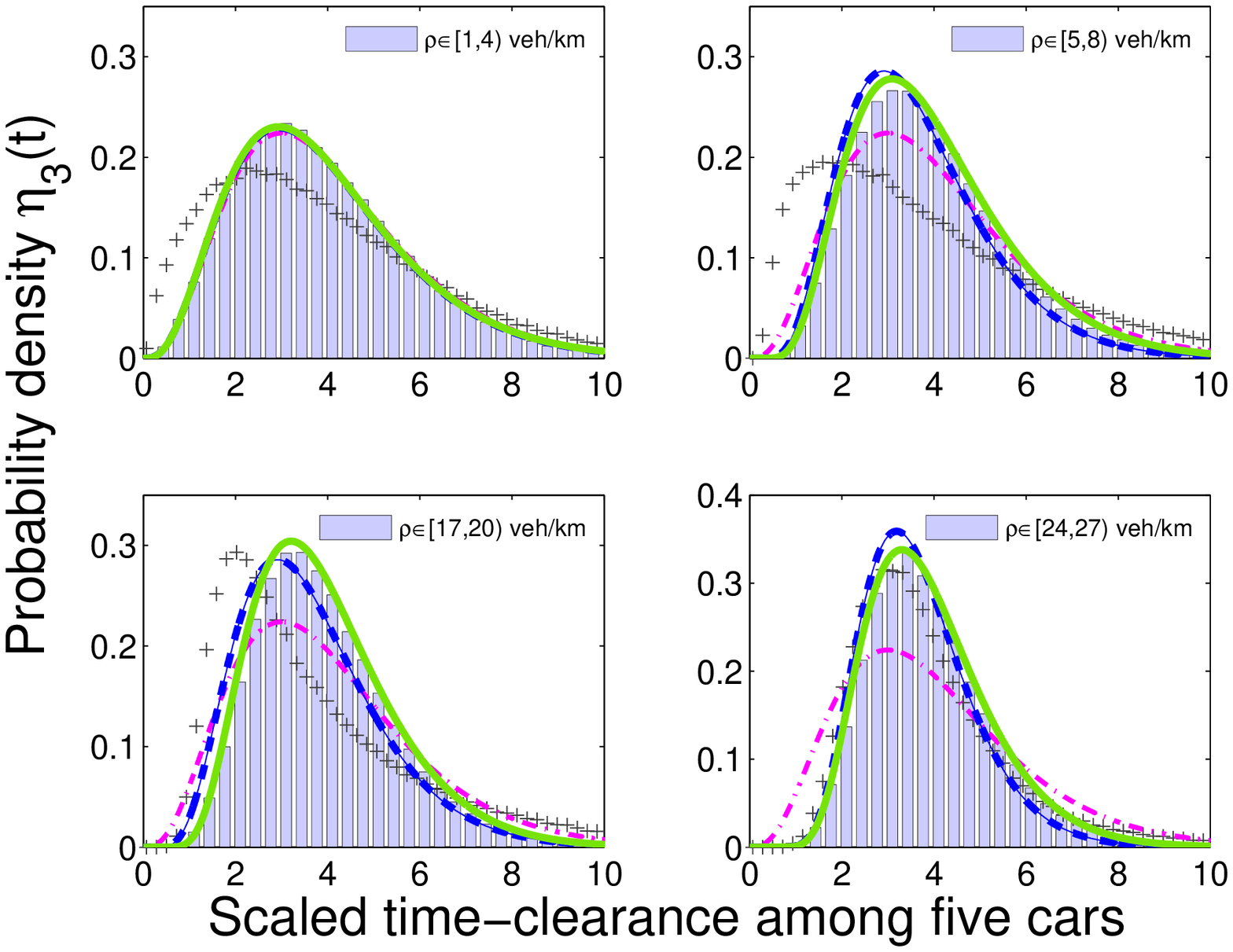}\\ \vspace*{0.5cm}
\hspace*{-2mm}\includegraphics[width=10cm, angle=0]{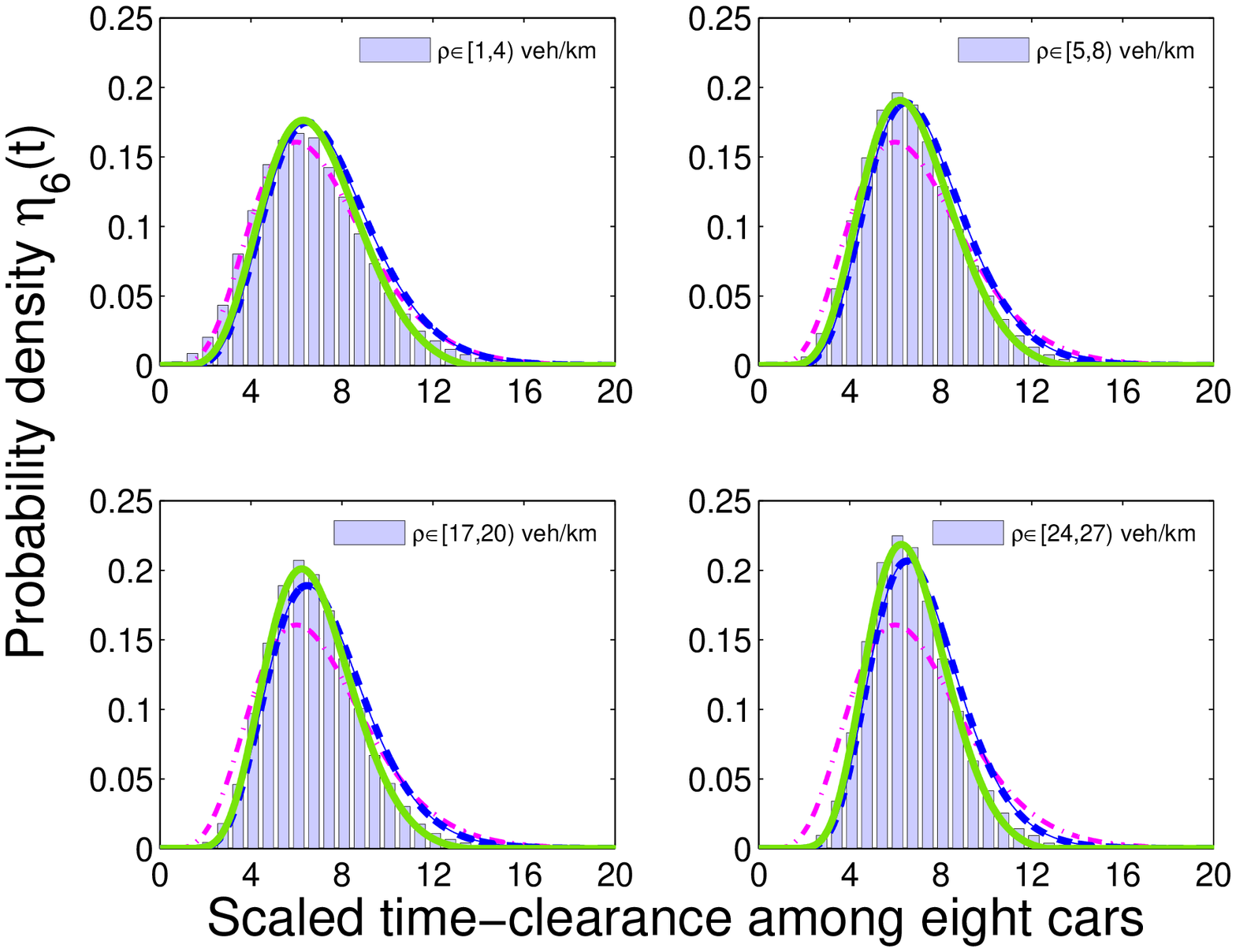}
\end{center}
\caption{Statistical distribution of time multi-clearances for low density regimes. The bars shows the empirical probability density for clear time-interval among five (or eight) succeeding vehicles moving in main lane (for various density regimes -- see legend for details). The blue dashed curves represent the prediction \eqref{nta-clearancka} plotted for $n=3$ (top plot) or $n=6$ (bottom plot), and for the fitted value of coefficient $\beta=\beta_{\fit}$ obtained by minimizing the weighted error-function \eqref{Miluna}. Continuous curves (green) display the analytical approximation \eqref{TC-sigma-aproximace-to-fit} plotted for the fitted values of $\beta=\beta_{\fit}$ and $\epsilon=\epsilon_{\fit}$ specified by minimizing the weighted error-function \eqref{Misule}. For clearness, we also plot the dash-dotted curves (magenta) visualizing the multi-clearance distribution \eqref{multi-poisson} valid for an occurrence of independent events. Plus signs illustrates how the time multi-clearances (gauged by the fast-lane-detectors) differ from those detected in the main lane.   \label{fig:multiclearance_ctyrobrazek_free}}
\end{figure}
\begin{figure}[htbp]
\begin{center}
\hspace*{-2mm}\includegraphics[width=10cm, angle=0]{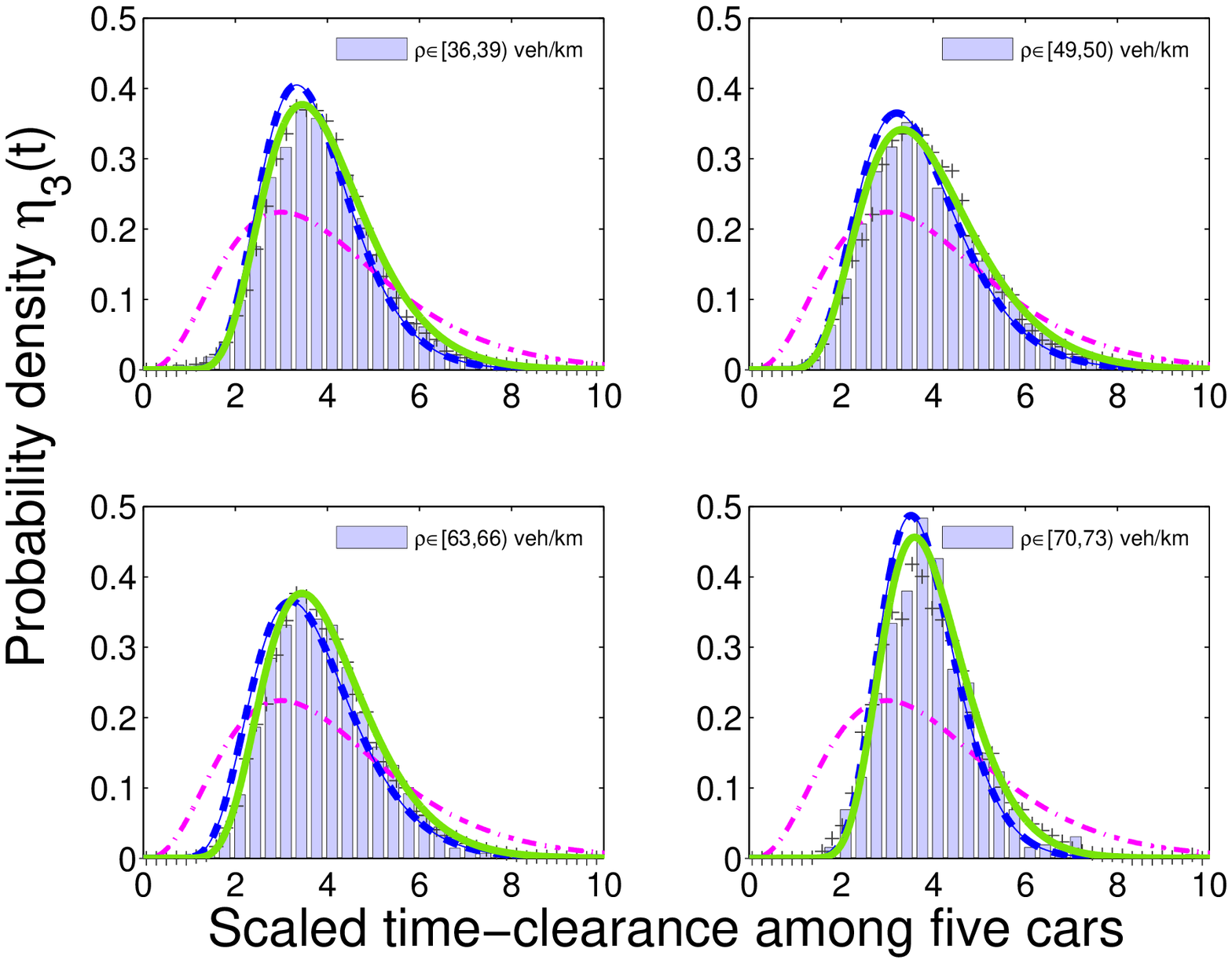}\\ \vspace*{0.5cm}
\hspace*{-2mm}\includegraphics[width=10cm, angle=0]{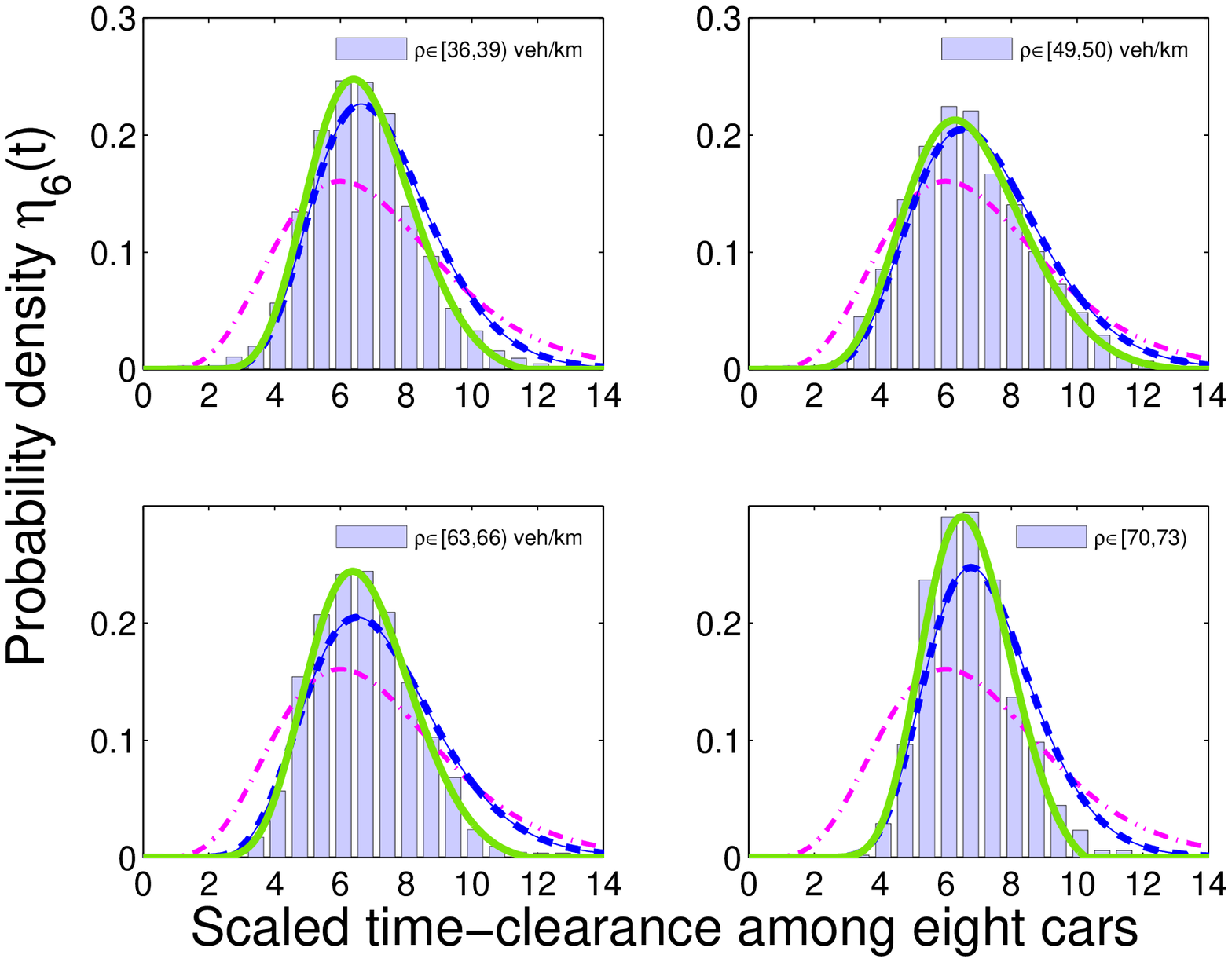}
\end{center}
\caption{Statistical distribution of time multi-clearances for high density regimes. The bars shows the empirical probability density for clear time-interval among five (or eight) succeeding vehicles moving in main lane (for various density regimes -- see legend for details). The blue dashed curves represent the prediction \eqref{nta-clearancka} plotted for $n=3$ (top plot) or $n=6$ (bottom plot), and for the fitted value of coefficient $\beta=\beta_{\fit}$ obtained by minimizing the weighted error-function \eqref{Miluna}. Continuous curves (green) display the analytical approximation \eqref{TC-sigma-aproximace-to-fit} plotted for the fitted values of $\beta=\beta_{\fit}$ and $\epsilon=\epsilon_{\fit}$ specified by minimizing the weighted error-function \eqref{Misule}. For clearness, we also plot the dash-dotted curves (magenta) visualizing the multi-clearance distribution \eqref{multi-poisson} valid for an occurrence of independent events. Plus signs illustrates how the time multi-clearances (gauged by the fast-lane-detectors) differ from those detected in the main lane.       \label{fig:multiclearance_ctyrobrazek_congested}}
\end{figure}

To be factual, we divide the entire interval of traffic densities into the subintervals $[0,3),$ $[1,4),$ $[2,5),$ and so on and analyze the freeway multi-clearance distribution $p(\varkappa)$ separately in each subinterval. Similar approach has been examined in the articles (\cite{Helbing_and_Krbalek, Krbalek_gas, Kesting, Traffic_NV}) and reflects the known fact that the headway distributions are substantially influenced by the changing location of the traffic ensemble in the phase diagram. To prevent the mixing of the states with different vigilance of car drivers (or with different temperature of associated heat bath -- in the thermodynamic interpretation of vehicular traffic), we inspect the empirical multi-clearances separately in each density interval.  Such an approach permits the investigation of density-dependent evolution of multi-clearance distributions.\\

In the Fig. \ref{fig:multiclearance_ctyrobrazek_free} and \ref{fig:multiclearance_ctyrobrazek_congested} there are plotted the empirical multi-clearance distributions (histograms) against the poissonian distribution \eqref{multi-poisson}, zero approximation \eqref{nta-clearancka} calculated for $n=3$ or $n=5,$ and the final analytical prediction
\BE \eta_n^{(\epsilon)}(t) = \Theta(t)\frac{A^{n+1}}{n!} t^n \e^{-(n+1)^2\beta/t}\e^{-Dt}\bigl(1-\epsilon t\bigr). \label{TC-sigma-aproximace-to-fit}\EE
The values of the fitted  parameters ($\beta$ in \eqref{nta-clearancka}, and $\beta, ~ \epsilon$ in \eqref{TC-sigma-aproximace-to-fit}) have been determined by means of formulae
\BE \beta_{\fit}^{\bullet} = \argmin_{\beta\in[0,\infty)} \int_0^\infty \bigl|\eta_n(t)-p(t)\bigr|^2 t\e^{-t/4}\,\dt, \label{Miluna} \EE
\BE \bigl(\beta_{\fit},\,\epsilon_{\fit}\bigr) = \argmin_{\beta\in[0,\infty),\epsilon\in[0,\infty)} \int_0^\infty \bigl|\eta_n^{(\epsilon)}(t) -p(t)\bigr|^2 t\e^{-t/4}\,\dt, \label{Misule}\EE
i.e. minimizing the statistical distance $\chi_n(\beta,\epsilon)=\int_0^\infty |f_n(t)-p(t)|^2 t\e^{-t/4}\,\dt$ (weighted by the factor $t\e^{-t/4}$) cumulating the weighted deviations between theoretical prediction $f_n(t)$ and empirical frequency $p(t).$ \\

\begin{figure}[htbp]
\begin{center}
\hspace*{-2mm}\includegraphics[width=7.5cm, angle=0]{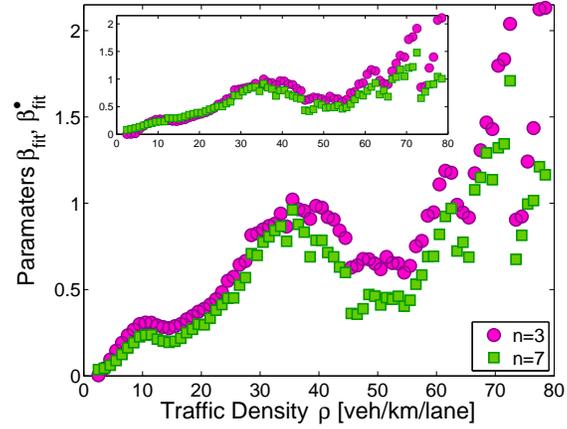}
\end{center}
\caption{Optimal values of the parameter $\beta$ in theoretical distributions. Signs (circles for $n=3$ and squares for $n=7$) visualize the value of the fitted parameter $\beta_{\fit}$ (see the relation \eqref{Misule}) that minimizes the statistical distance between theoretical curve \eqref{TC-sigma-aproximace-to-fit} and empirical histogram $p(t)$ (enumerated for main-lane data only). In the inset there is plotted the value of the thermal parameter $\beta_{\fit}^{\bullet}$ considered in \eqref{Miluna}.
\label{fig:parameters_beta}}
\end{figure}

\begin{figure}[htbp]
\begin{center}
\hspace*{-2mm}\includegraphics[width=7.5cm, angle=0]{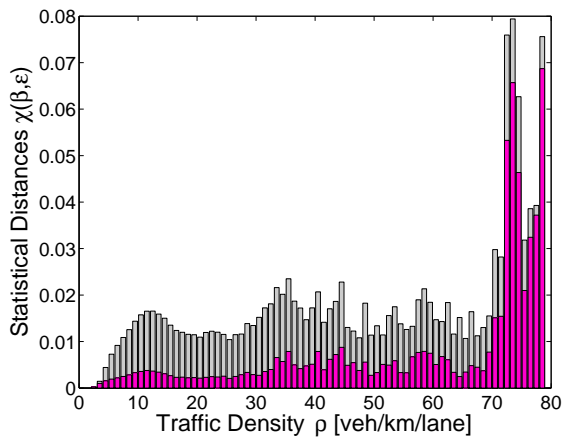}
\end{center}
\caption{Statistical distance between theoretical and empirical distributions. We plot the values of statistical distances $\chi_n(\beta_{\fit}^{\bullet})$ and $\chi_n(\beta_{\fit},\epsilon_{\fit})$ between main-lane multi-clearance distribution $p(t)$ and zero-approximation \eqref{nta-clearancka} (pale bars) or final formula \eqref{TC-sigma-aproximace-to-fit} (dark bars), respectively.   \label{fig:statistical_distances}}
\end{figure}

As visible in the Fig. \ref{fig:multiclearance_ctyrobrazek_free} and \ref{fig:multiclearance_ctyrobrazek_congested} the freeway multi-clearance distribution delineated for low-density states coincides with the distribution \eqref{multi-poisson}, which confirms the surmise that vehicles in free traffic regime are moving as independent elements. Thus, their statistics is purely poissonian. As the traffic density rises one can detect larger deviations from \eqref{multi-poisson}, which demonstrates stronger interactions among the cars. Roughly speaking, the fitted thermal parameter $\beta$ is increasing (in both cases: zero-approximation and also the final probability distribution) with the traffic density (see the Fig. \ref{fig:parameters_beta}).  The regions of the temporal descent in $\beta$ value agree with the critical regions in the fundamental diagram. To be precise, the transmission between traffic regimes (from free to congested regime) causes the transient consolidation of traffic, which leads to a reduction of driver's mental-strain. Since (see the \cite{Traffic_NV}) the thermal parameter reflects a level of mental pressure, the detected drop in course of $\beta=\beta(\varrho)$ is expectable. We add that the evolution of thermal parameter $\beta$ corresponds to the behavior of the associated quantity investigated within the scope of the articles \cite{Helbing_and_Krbalek,Krbalek_gas,Traffic_NV,KRB-Kybernetika}. Moreover, in all traffic states the distribution \eqref{TC-sigma-aproximace-to-fit} fits the real-road data more impressively than the original approximation \eqref{nta-clearancka}, as comprehensible. Such a fact is clearly visible in the Fig. \ref{fig:statistical_distances} where the statistical distance $\chi(\beta,\epsilon)$ between theoretical and empirical distributions are outlined.\\

\begin{figure}[htbp]
\begin{center}
\hspace*{-2mm}\includegraphics[width=7.5cm, angle=0]{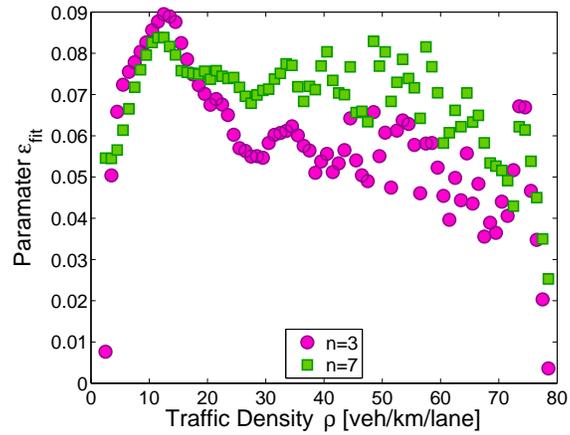}
\end{center}
\caption{Optimal value of the parameter $\epsilon$ in the theoretical distribution \eqref{TC-sigma-aproximace-to-fit}. Signs (circles for $n=3$ and squares for $n=7$) visualize the value of the fitted parameter $\epsilon_{\fit}$ (see the relation \eqref{Misule}) that minimizes the statistical distance between theoretical curve \eqref{TC-sigma-aproximace-to-fit} and empirical histogram $p(t)$ (enumerated for main-lane data only). \label{fig:parameters_epsilon}}
\end{figure}

The important component of our research is a discussion about stability of the above-introduced fitted procedure, i.e. stability of the detected interaction-parameter $\beta.$ The explorations of the dependency between the estimated interaction-parameter $\beta_{\fit}$ and number $n$ of cumulated clearances shows only insignificant fluctuations in $\beta_{\fit}=\beta_{\fit}(n).$ Such a fact is illustrated comprehensibly in the Fig. \ref{fig:parameters_beta}. It means that the value of $\beta_{\fit}$ is only slightly influenced by the order of the multi-clearance distribution $\eta_n^{(\epsilon)}(t).$ Therefore the estimated interaction-parameter $\beta_{\fit}$ represents a alternative traffic quantity whose values reflect the traffic state.

\section{Three-phases traffic -- evolution of multi-clearance distribution}\label{sec:sesta}

Finally, we try to connect the presented results with the three-phase traffic theory (see \cite{Review-Kerner}). The aim is to explore the essence of changes in multi-clearance distributions and detect a possible phase changes in the phase surface $\rho \times J.$ For these purposes we divide the entire phase space into the separate two-dimensional subregions $[\rho,\rho+\Delta\rho)\times [J,J+\Delta J).$ By analogy to the previous approach we investigate empirical distributions separately in each subregion. Surprisingly, in spite of the existing opinion, in all traffic phases the multi-clearances belong to the same family of distributions \eqref{TC-sigma-aproximace}. However, the resulting value $\beta^{\bullet\bullet}_{\fit}$ (obtained with help of the formula \eqref{Misule}) depends on density and flow, i.e. $\beta^{\bullet\bullet}_{\fit}=\beta^{\bullet\bullet}_{\fit}(\rho,J).$\\

\begin{figure}[htbp]
\begin{center}
\hspace*{-2mm}\includegraphics[width=9cm, angle=0]{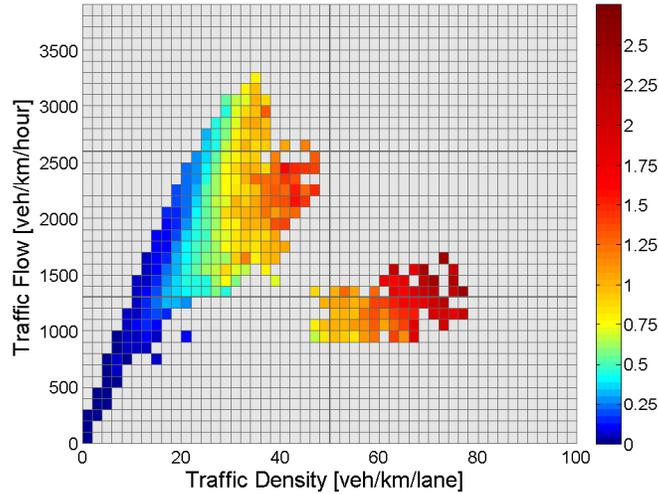}
\end{center}
\caption{Thermal parameter as a function of traffic density and traffic flux. We plot the dependency of the thermal parameter $\beta^{\bullet\bullet}_{\fit}$ occurring in the formula \eqref{Misule} for various subregions in the phase diagram $\rho \times J.$   \label{fig:temperature_3D}}
\end{figure}

As comprehensible from the Fig. \ref{fig:temperature_3D} the basic trends in the partial dependency $\beta^{\bullet\bullet}_{\fit}=\beta^{\bullet\bullet}_{\fit}(\rho)$ fully corresponds to the behavior $\beta_{\fit}=\beta_{\fit}(\rho)$ obtained in the section \ref{sec:pata}. Moreover, there is distinctly visible how the thermal parameter is adapted to the actual position in the phase space. The phase transition of the traffic system from free flows to synchronized flows (and vice versa) is accompanied by the corresponding changes in $\beta.$ Note that in the transitional region (between 35 and 55 $veh/km/lane$) the stable growth of thermal-like coefficient $\beta$ is temporarily attenuated (in fact, $\beta$ falls), which endorses the hypothesis on phase transition. Also, there is transparently demonstrated how the different traffic phases (i.e. drivers under a different mental pressure) influence the statistics of vehicular interactions.\\

For completeness, we emphasize again that all the statistical test discussed in this article have been executed for the main lane traffic data only. As demonstrated in the Fig. \ref{fig:multiclearance_ctyrobrazek_free} and \ref{fig:multiclearance_ctyrobrazek_congested} the time multi-clearance distributions detected in the fast-lane data show the well-known discrepancies with the main-lane data (predominantly in the regions of low densities). For denser streams the fast-lane distributions converge to the main-lane distributions, which endorses the belief of scientists that in congested traffic states the correlations among the vehicles in different traffic lanes are much stronger than in free-flow regimes.

\section{Summary, conclusion, and future prospects}\label{sec:Summary}

To conclude, this article deals with cumulative time clearances among several subsequent vehicles passing a given point of an expressway. More specifically, we are concentrated predominantly on time intervals between two following occurrences: 1. the rear bumper of $k$th car has intersected the detector line, 2. the front bumper of $(k+n)$th car has intersected the detector line. Furthermore, we subtract the time-periods when vehicles are occupying the detectors, which means that a fundamental quantity for this research is the time multi-clearance. Using the local thermodynamical traffic model (originally introduced in \cite{Helbing_and_Krbalek} and solved analytically in \cite{Krbalek_gas}) with short-range repulsive potential among the elements we have derived (applying theory of functional convolutions) a mathematical formula for time multi-clearance distribution. This final theoretical distribution represents an one-parametric family of functions, where the one and only fitting parameter $\beta$ reflects the relevant flow and density. The obtained analytical predictions have been successfully compared with statistics of freeway data. The detected correspondence between theoretical and empirical distributions allows an elaborated insight into changes of vehicular-traffic microstructure influenced by the momentary traffic state.\\

In spite of the existing conviction that time-headway distributions in the different phases are markedly different, the stochastic analysis of multi-clearances reveals a different view: the statistical distribution of traffic clearances (and multi-clearances) in all traffic phases belongs to the same one-parametric family of distributions. Those distributions are varying from the Poisson distribution (detected for low-density states) to the low-variance distribution \eqref{TC-sigma-aproximace-to-fit} that ascertains  a presence of stronger correlations among neighboring vehicles, i.e. stronger vehicular synchronization than in free-flow states. The existing conviction on two types of traffic headway-distributions is probably a consequence of insufficient sorting of freeway data. In the earlier scientific papers (dealing with traffic headways) authors usually separate vehicular data into two/three parts only (according a traffic phase) and in fact they combine all the congested states into a single ensemble. But, as apparent from this research, the congested-phase-distributions are significantly varying according to the traffic flow and density. For that reason the binary data-separation seems to be deficient. \\

Another substantial outcome of these considerations is the fact that the inter\-action-pa\-ra\-meter $\beta$ (quantifying a measure of a vehicular synchronization) represents an alternative traffic quantity which reflects an actual traffic state. Moreover, the value $\beta$ is (in view of the fact that all clearances are scaled) directly connected to the statistical variance of traffic clearances, and its estimated value can be therefore obtained by a simple comparison of the empirical and theoretical variances, which is quite effortless. \\

However, the open problem remains how to approximate the multi-clear\-ance distribution for free-flow vehicles moving in fast lanes. Presence of large amount of vehicular \emph{leaders} in free-traffic regimes, i.e. comparable percentage of leaders and \emph{followers}, causes that the clearance distribution is probably a compound of two partial distributions (first one for leaders, second one for followers). Thus, the expected approach leading to analytical predictions for fast-lane clearances can be found in the theory of finite mixture distributions (similarly to semi-poissonian model discussed in \cite{Buckley}).

\subsection*{Acknowledgement}
The author would like to thank Cecile Appert-Rolland (Laboratoire de Phy\-si\-que Th\'eorique
Universit\'e de Paris-Sud, Orsay) for valuable remarks which have been conductive to the presented research. This work was supported by the Ministry of Education, Youth and Sports of the Czech Republic within the project MSM 6840770039 and by the Czech Technical University within the project SGS12/197/OHK4/3T/14.

\bibliographystyle{elsarticle-num}
\bibliography{<your-bib-database>}

\end{document}